\documentclass[prb,twocolumn, superscriptaddress]{revtex4}
\usepackage{epsfig}
\usepackage{graphicx}
\usepackage{dcolumn}
\usepackage{bm}
\usepackage[english]{babel}
\usepackage{amsmath, amssymb}
\usepackage{pdfpages}

\begin{document}
	
\title{Remarkable nuances of crystallization: \\ From ordinary crystal nucleation to rival mechanisms of crystallite coalescence}

\author{Bulat N. Galimzyanov}
\email{bulatgnmail@gmail.com}
\affiliation{Kazan Federal University, 420008 Kazan, Russia}
\affiliation{Udmurt Federal Research Center of the Ural Branch of the Russian Academy of Sciences, 426067 Izhevsk, Russia}

\author{Vladimir I. Ladyanov}
\affiliation{Udmurt Federal Research Center of the Ural Branch of the Russian Academy of Sciences, 426067 Izhevsk, Russia}

\author{Anatolii V. Mokshin}
\email{anatolii.mokshin@mail.ru}
\affiliation{Kazan Federal University, 420008 Kazan, Russia}
\affiliation{Udmurt Federal Research Center of the Ural Branch of the Russian Academy of Sciences, 426067 Izhevsk, Russia}

\begin{abstract}
Crystal growth and crystal coalescence processes in supercooled systems strongly depend on the concentration of crystallization centers. We perform atomistic dynamics simulations of the crystallization process in the ultrathin metallic film at different supercooling levels corresponding to supercooled liquid and amorphous solid states. Scaled relations are applied to identify the characteristic regimes in the time-dependent crystalline nuclei concentration: steady-state nucleation regime, saturation regime and coalescence regime. We show that the crystal growth at the saturation regime appears due to mixing nucleation and coalescence processes. We find that the crystallite coalescence realizes mainly through the mechanism of restructurization/absorption of crystal nuclei, whereas the mechanism of oriented attachment is manifested only at low levels of supercooling.
\end{abstract}
	
\maketitle

\section{Introduction}

Crystal growth is complex process of formation of an ordered stable phase~\cite{Kashchiev_2000,Sosso_Chen_2016,Schmelzer_Abyzov_2018}. According to classical models~\cite{Kelton_Greer_2010,Turnbull_Fisher_1949,Weinberg_Poisl_2002,Mokshin_Galimzyanov_2017,Galenko_2018}, crystalline nuclei (stable crystallization centers) grow by attachment of particles (atoms, molecules, ions) to their surface. This scenario is usually implemented at an early stage of the crystallization process, when there is no a direct contact between the growing nuclei due to the low concentration of these nuclei. Coalescence of two or more nuclei and subsequent formation of a new larger crystallite is another scenario of crystal growth, which occurs usually at large concentrations of the nuclei~\cite{Zhang_2010,Ivanov_Fedorov_2014,Niederberger_2006}.

Understanding the nuclei coalescence mechanisms is of great importance from the viewpoint of new functional materials production~\cite{Ivanov_Fedorov_2014,Grammatikopoulos_Kioseoglou_2019}. One of the most intriguing is crystal growth through a mechanism of oriented attachment. This mechanism is manifested in the crystallizing low-viscous supercooled liquids, and is detectable experimentally using high-resolution transmission electron microscopy~\cite{Fedorov_Osiko_2014,Li_Nielsen_2012,Zhang_Penn_2014}. According to this mechanism, coalescence of crystalline nuclei arises as a result of their mutual rotation and the subsequent decrease of their crystallographic misorientation~\cite{Ivanov_Fedorov_2014,Zhu_Liang_2018}. It was reported in Refs.~\cite{Yoo_Lee_2013,Kracker_Zscheckel_2019}, that the crystal growth through this mechanism was also found in crystallizing high-viscous systems such as glass-ceramics. Nevertheless, as was found, due to the very slow particle dynamics in viscous systems, the mechanism of oriented attachment is not decisive for crystal growth.

The present work is devoted to study the growth of crystalline nuclei in a model liquid at different levels of supercooling. The main attention is paid to reveal the role of the mechanism of oriented attachment in the crystal growth process.

\section{About the system}

We perform atomistic dynamics simulations of the crystallization process in an ultrathin metallic film. The simulated system is composed of $N=14\,700$ particles located inside a cell of the volume $V=L_{x}\times L_{y}\times L_{z}$ with $L_{x}=L_{y}\simeq11.5L_{z}$ and $L_{z}\simeq4.8\,\sigma$~\cite{Galimzyanov_Mokshin_2019}. The particles interact through the short-ranged oscillating Dzugutov potential~\cite{Dzugutov_1992}. Supercooling states are achieved by fast cooling of well equilibrated liquid sample to temperatures from the range $T\in [0.5;\,1.4]\,\epsilon/k_{B}$ at the pressure $p=15\,\epsilon/\sigma^{3}$. For this isobar, the glass transition temperature and the melting temperature are $T_{g}=0.78\,\epsilon/k_{B}$ and $T_{m}=1.72\,\epsilon/k_{B}$, respectively. Cluster analysis and characterization of the local structure of the crystallizing system are performed by computing the local orientational order parameters~\cite{Mickel_Kapfer_2013}. Details of the simulation protocol are given in Ref.~\cite{Galimzyanov_Mokshin_2019}.
\begin{figure*}
	\centering
	\includegraphics[width=1.0\linewidth]{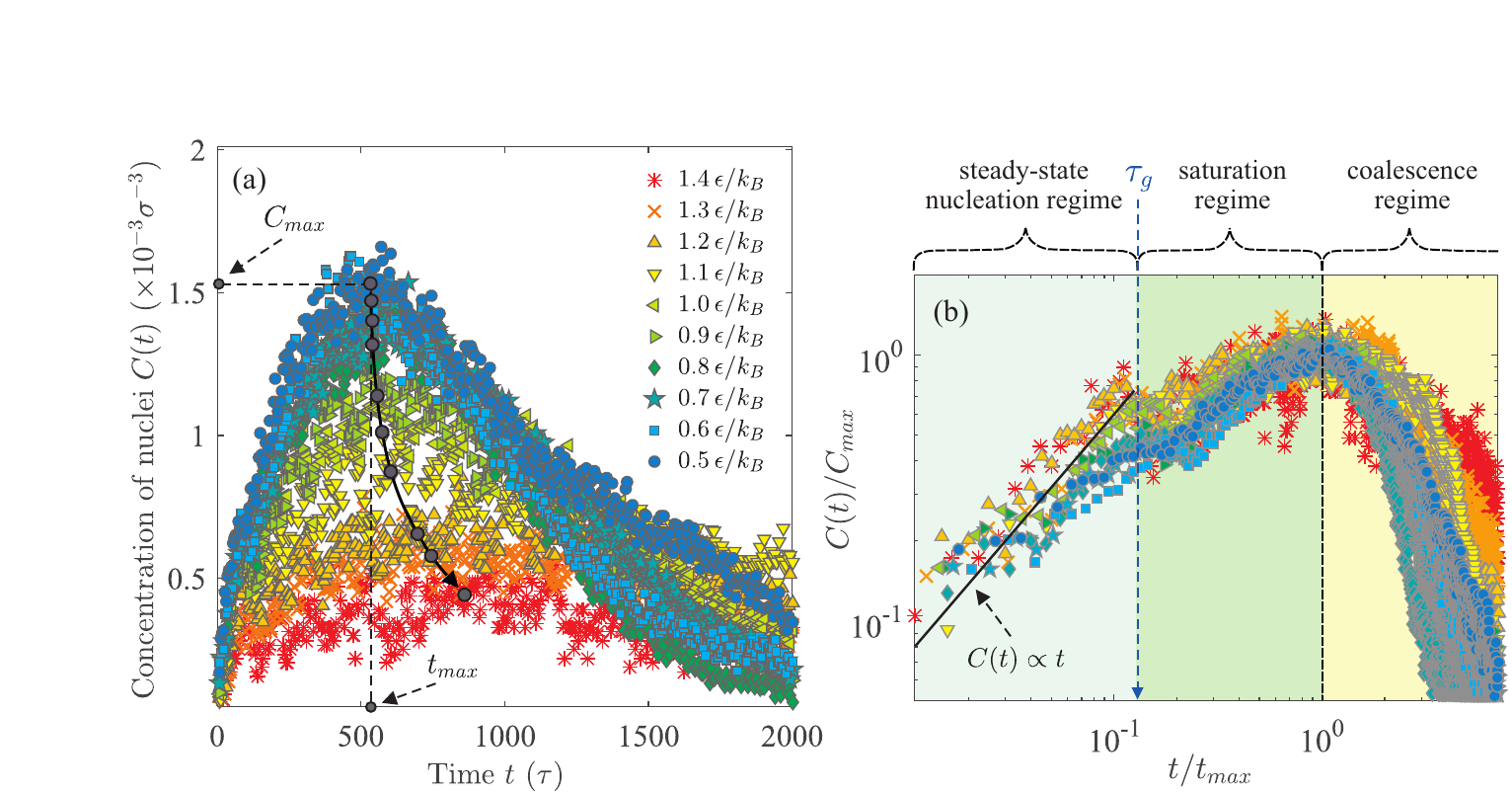}
	\caption{(a) Time-dependent concentration $C(t)$ of supercritical crystalline nuclei at various temperatures. (b) Scaled concentration $C(t)/C_{max}$ as function of the reduced time $t/t_{max}$ in double logarithmic scale. The dashed lines separate the different regimes in $\log[C(t)/C_{max}]$ versus $\log[t/t_{max}]$ plot. The quantity $\tau_{g}\simeq(0.1\pm0.01)\, t_{max}$ is the waiting time for a first coalescence event.}\label{fig_1}
\end{figure*}

In the present work, all the quantities are expressed in terms of the parameters of the potential: the effective particle diameter $\sigma$ and the energy parameter $\epsilon$. The time unit is $\tau=\sigma\sqrt{m/\epsilon}$, where $m$ is the particle mass; the temperature unit is $\epsilon/k_{B}$, where $k_{B}$ is the Boltzmann constant; the concentration unit is $\sigma^{-3}$.

\section{Results and discussion}

For considered thermodynamic ($p$,~$T$)-states, the processes of crystal nucleation and crystal growth in the system under consideration are monitored over the time interval $t\simeq10^{4}\,\tau$. For elementary metallic systems, this time interval is of the order of tens nanoseconds~\cite{Belchior_2005}, that is evaluated easily with the known parameters $\sigma$ and $\epsilon$ of potentials and the mass $m$ for these systems. Recently, we have shown that the crystallization process in the considered system starts with formation of the (critically-sized) crystalline nuclei, the size of which varies from $n_{c}\simeq40$ particles (at the temperature $T=0.5\,\epsilon/k_{B}$) to $n_{c}\simeq100$ particles (at the temperature $T=1.4\,\epsilon/k_{B}$)~\cite{Galimzyanov_Mokshin_2019}.  In this study, we will consider the growth of the supercritical crystalline nuclei, which are the nuclei of the size $n$ larger than the critical size $n_c$.

Figure~\ref{fig_1}a shows the concentration $C(t)$ of supercritical nuclei computed for the crystallizing samples at different temperatures. It is important to note that only supercritical nuclei are taken in account to evaluate the quantity $C(t)$~\cite{Galimzyanov_Mokshin_2019}. As seen from Figure~\ref{fig_1}a, for the ($p,T$)-states with all the considered temperatures, the time-dependent concentration $C(t)$ demonstrates the common scenario: the number of nuclei increases rapidly, reaches a maximum value and then starts to decrease. As expected, with increase of the temperature $T$ (i.e. with decrease of the supercooling level of the system), the maximum value of $C(t)$ is achieved later. We denote location of the maximum of the function $C(t)$ on the time scale as $t_{max}$, whereas $C_{max} = C(t=t_{max})$.

\begin{figure*}
	\centering
	\includegraphics[width=1.0\linewidth]{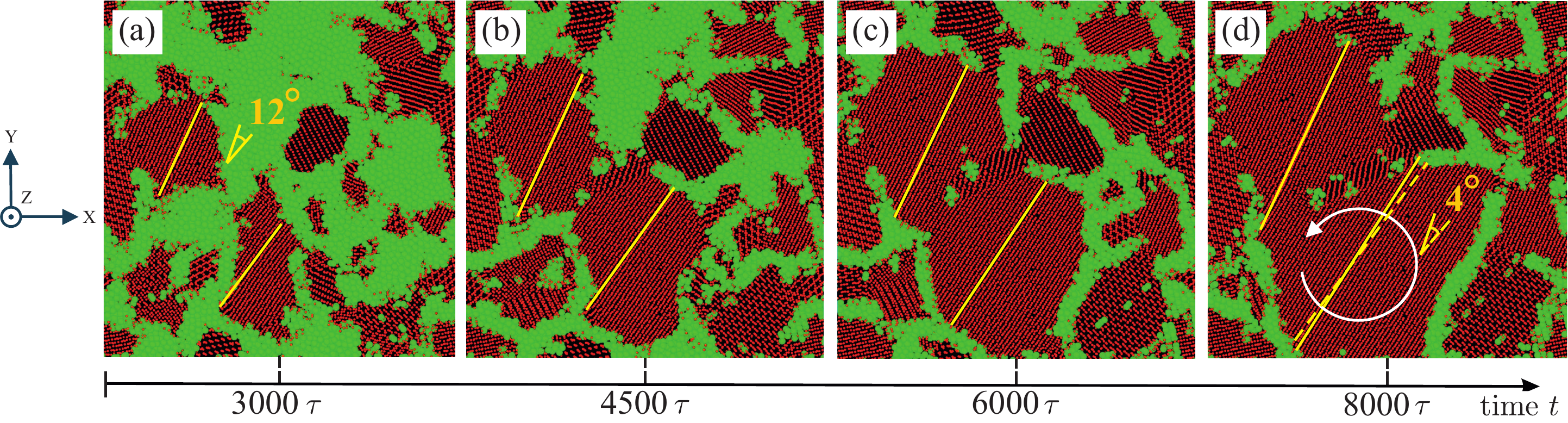}
	\caption{Configurations of the system crystallizing at the temperature $T=1.2\,\epsilon/k_{B}$ at different times: (a) $3000\,\tau$, (b) $4500\,\tau$, (c) $6000\,\tau$ and (d) $8000\,\tau$. The solid lines indicate crystallographic directions for two crystallites, which will coalesce through the  mechanism of oriented attachment. The misorientation angle, which accounts for the difference between these directions, is about $12$ degrees at the time $t=3000\,\tau$. During time from the moment $t=3000\,\tau$ to $8000\,\tau$, the angle reduces to value $\simeq4$ degrees.}\label{fig_2}
\end{figure*}

To compare the time-dependent concentrations at different temperatures we present these data in a rescaled form, namely, $C(t)/C_{max}$ vs. $t/t_{max}$ (see Figure~\ref{fig_1}b). As seen, three regimes can be clearly recognized. The first regime corresponds to the times up to $t/t_{max} \sim 0.1$. The linear dependence $C(t)\propto t$ occurs for the regime, that is well-known scenario for the \textit{steady-state nucleation regime}~\cite{Kashchiev_2000}. Here, there is a spontaneous increase the number of crystalline nuclei, which demonstrate resistant growth. The growth of these nuclei usually occurs by attachment of particles to their surface. In time, the concentration of the growing nuclei becomes so high that the nuclei begin to come in contact with each other. The second regime denoted as the \textit{saturation regime} is started with the first event of the nuclei coalescence, that occurs at time $\tau_{g} \simeq (0.1 \pm 0.02) t_{max}$ for the system at all the considered crystallization temperatures. At this regime, the structural ordering proceeds in a mixed manner through the crystal nucleation and the crystal coalescence, although the nucleation process prevails. The maximum of $C(t)$ reveals the start of the next regime -- the \textit{coalescence regime}, where the coalescence process is of the main importance in the structural ordering.

\begin{figure*}
	\centering
	\includegraphics[width=1.0\linewidth]{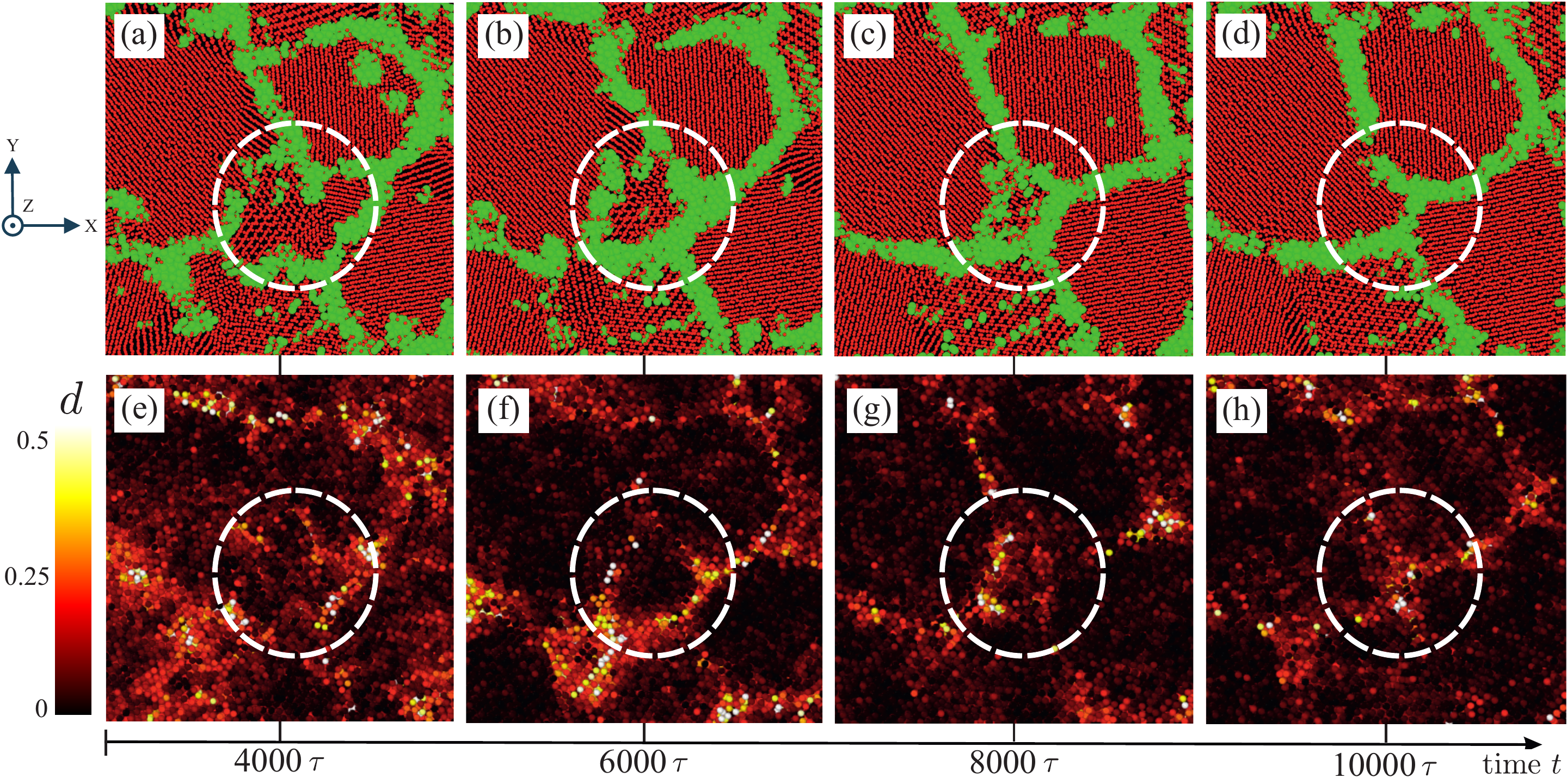}
	\caption{Snapshots of the system at different times and at the temperature $T=0.6\,\epsilon/k_{B}$. In lower panels, the color of the particles depends on value of the parameter $d$ calculated by Eq.~(\ref{eq_mobility_parameter}): the higher value of the parameter $d$ for a particle, the closer the color to white. The dashed circle denotes the collapsing nucleus.}\label{fig_3}
\end{figure*}

Decrease of the concentration $C(t)$ with time is the distinctive feature of the coalescence regime. We found that there are two mechanisms for nuclei coalescence, which differ from each other: the oriented attachment mechanism and the mechanism of the full absorption of the crystal nuclei by other growing crystal nuclei. 

The events, where the nuclei coalescence proceeds through the mechanism of oriented attachment, are detected for the system at low supercooling levels (i.e. at the temperatures $T>1.0\,\epsilon/k_{B}$). As an example, we show in Figures~\ref{fig_2} the coalescence of the two crystallites as a process that develops over time. Initially, the angle, which accounts for the difference between the crystallographic directions in these crystallites, is about $12$ degrees [Figure~\ref{fig_2}(a)]. The small difference in the mutual crystallographic orientation of the crystallites as well as their small size favor to coalescence process through this mechanism.  However, the rotational movement of these crystallites is extremely slow [Figures.~\ref{fig_2}(a-d)], that is due to high viscosity and/or high density~\cite{Galimzyanov_Mokshin_2019}. As seen from Figures~\ref{fig_2}, during the time period $4000~\tau$ (from the moment $t=3000\,\tau$ to $8000\,\tau$), this misorientation angle reduces only to value $\simeq4$ degrees. We note that for extremely low density states, the mechanism of oriented coalescence is realized even when the initial angle is more than $40$ degrees~\cite{Cheng_Lian_2019}.

Restructurization and absorption of small crystallites is other mechanism rival to oriented coalescence.
This mechanism of the structural ordering is realized for the system at low as well as at deep levels of supercooling.
In Figures~\ref{fig_3}(a-d), we show the snapshots of the system crystallizing at the temperature~$T=0.6\,\epsilon/k_{B}$; here, each the snapshot  is appropriate to the concrete time moment. We note that the considered temperature state corresponds to a deep level of supercooling. The scenario of the structural ordering observed on this set of snapshots (the central part of the figures) is directly related to the mechanism of restructurization and absorption of a small nucleus. Since the process of crystal nuclei destruction should be accompanied by an increase of the mobility of the particles, then it is convenient to consider the parameter
\begin{equation}\label{eq_mobility_parameter}
d_{i}(t)=\frac{\left\langle[\vec{r}_{i}(t+\tau_{w})-\vec{r}_{i}(t)]^{2}\right\rangle}{\lambda^{2}},
\end{equation}
which provides an estimate for the relative mean displacement for an each particle over the time window $\tau_w = 10 \tau$ starting from the time moment $t$. Here, $\vec{r}_{i}(t)$ is the trajectory of an $i$th particle during the time $\tau_w = 10 \tau$ starting from the time moment $t$. The quantity $\lambda=1/\sqrt[3]{\rho}$ is the mean free path, and $\rho$ is the numerical density of the system~\cite{Galimzyanov_Mokshin_2019}. The angle brackets $<...>$ mean the time averaging. The larger mobility of the particles, the larger value of the parameter $d$. As seen in Figures.~\ref{fig_3}(e-h), the mobility of particles is higher in the boundary regions, where the processes of the particle attachment and detachment are pronounced. Note that due to the low mobility of particles at very deep levels of supercooling, polycrystalline structures are formed as a result of ordinary crystal nuclei coalescence~\cite{Galimzyanov_Yarullin_2018}.

In conclusion, we note that the specific regimes of crystallization related with crystal nucleation and different scenarios of the crystal nuclei coalescence can be revealed by means of the quantity which accounts for the time-dependent concentration of crystalline nuclei. We have shown that these regimes are distinguishable both at low and at deep levels of supercooling. The coalescence of the growing crystal nuclei occurs by means of two rival mechanisms: the restructurization/absorption of  crystal nucleus and the oriented attachment. 

\section*{Acknowledgements}
\noindent This work is supported by the Russian Science Foundation (project 19-12-00022).

\bibliographystyle{unsrt}

\begin{thebibliography}{99}

\bibitem{Kashchiev_2000}
D.~Kashchiev, Nucleation: Basic theory with applications, Butterworth-Heinemann, Oxford, 2000.

\bibitem{Sosso_Chen_2016}
G.C.~Sosso, J.~Chen, S.J.~Cox, M.~Fitzner, Ph.~Pedevilla, A.~Zen, and A.~Michaelides, Crystal nucleation in liquids: Open questions and future challenges in molecular dynamics simulations, Chem. Rev. 116 (2016) 7078--7116. https://doi.org/10.1021/acs.chemrev.5b00744

\bibitem{Schmelzer_Abyzov_2018}
J.W.P.~Schmelzer, A.S.~Abyzov, Crystallization of glass-forming melts: New answers to old questions, J. Non-Cryst. Solids 501 (2018) 11--20. https://doi.org/10.1016/j.jnoncrysol.2017.11.047

\bibitem{Kelton_Greer_2010}
K.~F.~Kelton and A.~L.~Greer, Nucleation in condensed matter, Elsevier, Amsterdam, 2010.

\bibitem{Turnbull_Fisher_1949}
D.~Turnbull and J.C.~Fisher, Rate of nucleation in condensed systems, J. Chem. Phys. 17 (1949) 71--73. https://doi.org/10.1063/1.1747055

\bibitem{Weinberg_Poisl_2002}
M.C.~Weinberg, W.H.~Poisl and L.~Granasy, Crystal growth and classical nucleation theory, C. R. Chim. 5 (2002) 765--771. https://doi.org/10.1016/S1631-0748(02)01433-9

\bibitem{Mokshin_Galimzyanov_2017}
A.V.~Mokshin and B.N.~Galimzyanov, Kinetics of crystalline nuclei growth in
glassy systems, Phys. Chem. Chem. Phys. 19 (2017) 11340--11353. https://doi.org/10.1039/C7CP00879A

\bibitem{Galenko_2018}
P.K.~Galenko, I.G.~Nizovtseva, K.~Reuther, and M.~Rettenmayr, Kinetics of the Formation of a Disordered Crystal Structure during High-Speed Solidification, JETP 127 (2018) 107--114. https://doi.org/10.1134/S106377611807004X

\bibitem{Zhang_2010}
J.~Zhang, F.~Huang, and Z.~Lin, Progress of nanocrystalline growth kinetics based on oriented attachment, Nanoscale 2 (2010) 18--34. https://doi.org/10.1039/b9nr00047j

\bibitem{Ivanov_Fedorov_2014}
V.K.~Ivanov, P.P.~Fedorov, A.Ye.~Baranchikov, V.V.~Osiko, Oriented attachment of particles: 100 years of investigations of non-classical crystal growth, Russ. Chem. Rev. 83 (2014) 1204--1222. https://doi.org/10.1070/RCR4453

\bibitem{Niederberger_2006}
M.~Niederberger and H.~C\"{o}lfen, Oriented attachment and mesocrystals: Non-classical crystallization
mechanisms based on nanoparticle assembly, Phys. Chem. Chem. Phys. 8 (2006) 3271--3287. https://doi.org/10.1039/B604589H

\bibitem{Grammatikopoulos_Kioseoglou_2019}
P.~Grammatikopoulos, M.~Sowwan, and J.~Kioseoglou, Computational Modeling of Nanoparticle Coalescence, Adv. Theory Simul. 2 (2019) 1900013.  https://doi.org/10.1002/adts.201900013

\bibitem{Fedorov_Osiko_2014}
P.P.~Fedorov, V.V.~Osiko, S.V.~Kuznetsov, O.V.~Uvarov, M.N.~Mayakova, D.S.~Yasirkina, A.A.~Ovsyannikova, V.V.~Voronov, V.K.~Ivanov, Nucleation and growth of fluoride crystals by agglomeration
of the nanoparticles, J. Cryst. Growth 401 (2014) 63--66. http://dx.doi.org/10.1016/j.jcrysgro.2013.12.069

\bibitem{Li_Nielsen_2012}
D.~Li, M.H.~Nielsen, J.R.I.~Lee, C.~Frandsen, J.F.~Banfield, J.J.~De~Yore, Direction-specific interactions control crystal growth by oriented attachment, Science 336 (2012) 1014--1018. https://doi.org/10.1126/science.1219643

\bibitem{Zhang_Penn_2014}
H.~Zhang, R.L.~Penn, Z.~Linc and H.~C\"{o}lfen, Nanocrystal growth via oriented attachment,
CrystEngComm 16 (2014) 1407--1408. https://doi.org/10.1039/C4CE90001D

\bibitem{Zhu_Liang_2018}
C.~Zhu, S.~Liang, E.~Song, Y.~Zhou, W.~Wang, F.~Shan, Y.~Shi, C.~Hao, K.~Yin, T.~Zhang, J.~Liu, H.~Zheng and L.~Sun, In-situ liquid cell transmission electron microscopy investigation on oriented attachment of gold nanoparticles, Nature Communications 9 (2018) 1--7. https://doi.org/10.1038/s41467-018-02925-6

\bibitem{Yoo_Lee_2013}
M.-Y.~Yoo, J.-H.~Lee, H.-M.~Jeong, K.-S.~Lim, P.~Babu, Enhancement of photoluminescence and upconversion in Er-Yb codoped nanocrystalline glass-ceramics, Optical Materials 35 (2013) 1922--1926. https://doi.org/10.1016/j.optmat.2012.12.020

\bibitem{Kracker_Zscheckel_2019}
M.~Kracker, T.~Zscheckel, C.~Thieme, K.~Thieme, T.~H\"{o}cheb and C.~R\"{u}ssel, Morphology, topography, and crystal rotation during surface crystallization of BaO/SrO/ZnO/SiO$_{2}$ glass, CrystEngComm 21 (2019)
1320--1328. https://doi.org/10.1039/C8CE02057D

\bibitem{Dzugutov_1992}
M.~Dzugutov, Glass formation in a simple monatomic liquid with icosahedral inherent local order, Phys. Rev. A 46 (1992) R2984--R2987. https://doi.org/10.1103/PhysRevA.46.R2984

\bibitem{Mickel_Kapfer_2013}
W.~Mickel, S.C.~Kapfer, G.E.~Schr\"{o}der-Turk, K.~Mecke, Shortcomings of the bond orientational order parameters for the analysis of disordered particulate matter, J. Chem. Phys. 138 (2013) 044501. https://doi.org/10.1063/1.4774084

\bibitem{Galimzyanov_Mokshin_2019}
B.N.~Galimzyanov, D.T.~Yarullin, A.V.~Mokshin, Structure and morphology of crystalline nuclei arising in a
crystallizing liquid metallic film, Acta Materialia 169 (2019) 184--192. https://doi.org/10.1016/j.actamat.2019.03.009

\bibitem{Belchior_2005}
M.~B\"{o}y\"{u}kata, E.~Borges, J.P.~Braga, J.C.~Belchior, Size evolution of structures and energetics of iron clusters (Fe$_{n}$, $n\leq36$): Molecular dynamics studies using a Lennard-Jones type potential,
J. Alloys Compd. 403 (2005) 349--356. https://doi.org/10.1016/j.jallcom.2005.06.008

\bibitem{Cheng_Lian_2019}
F.~Cheng, L.~Lian, L.~Li, J.~Rao, C.~Li, T.~Qi, Z.~Zhang, J.~Zhang, and Y.~Gao, Hybrid growth modes of PbSe nanocrystals with oriented attachment and grain boundary migration, Adv. Sci. 6 (2019) 1802202. https://doi.org/10.1002/advs.201802202

\bibitem{Galimzyanov_Yarullin_2018}
B.N.~Galimzyanov,~D.T.~Yarullin, and A.V.~Mokshin, Change in the crystallization features of supercooled liquid metal with an increase in the supercooling level, JETP Letters 107 (2018) 629--634. https://doi.org/10.1134/S0021364018100089

\end{thebibliography}

\end{document}